\documentclass[journal]{IEEEtran}

\ifCLASSINFOpdf
\else
   \usepackage[dvips]{graphicx}
\fi
\usepackage{url}
\usepackage{amsmath}
\usepackage{caption}
\usepackage{graphicx}
\usepackage{multirow}
\usepackage[pagebackref=true,breaklinks=true,letterpaper=true,colorlinks,citecolor=blue,linkcolor=blue,bookmarks=false]{hyperref}
\usepackage{amsthm,cite,url,graphicx,booktabs,lipsum,color,bm,caption,subcaption,soul}
\usepackage{pifont,tikz,paralist,multirow,amssymb}

\begin{document}

\title{AgentPolyp: Accurate Polyp Segmentation via Image Enhancement Agent}

\author{Pu Wang, Zhihua Zhang, Dianjie Lu, Guijuan Zhang, Youshan Zhang, Zhuoran Zheng
%\IEEEmembership{Member, IEEE}
\thanks{Corresponding author: Zhuoran Zheng}
\thanks{Pu Wang and Zhihua Zhang is with the School of Mathematics, Shandong University,  Jinan 250100 , China, (e-mail: 202411943@mail.sdu.edu.cn, zhangzhihua@sdu.edu.cn).}%This paragraph of the first footnote will contain the date on which you submitted your paper for review. It will also contain support information, including sponsor and financial support acknowledgment. For example, ``This work was supported in part by the U.S. Department of Commerce under Grant BS123456.'' 
\thanks{Dianjie Lu and Guijuan Zhang is with the School of Information Science and Engineering, Shandong Normal University,  Jinan 250100 , China, (e-mail: ludianjie@sdnu.edu.cn, zhangguijuan@sdnu.edu.cn).}
\thanks{Youshan Zhang is with the Department of Artificial Intelligence and Computer Science,  Yeshiva University, New York, NY, US (e-mail: yz945@cornell.edu).}
\thanks{Zhuoran Zheng is with the School of cyber science and technology, Sun Yat-sen University, Shenzhen 518000, China, (e-mail: zhengzr@njust.edu.cn).}}%The next few paragraphs should contain the authors' current affiliations, including current address and e-mail. For example, F. A. Author is with the National Institute of Standards and Technology, Boulder, CO 80305 USA (e-mail: author@boulder.nist.gov).

\markboth{Journal of \LaTeX\ Class Files, Vol. 14, No. 8, August 2015}
{Shell \MakeLowercase{\textit{et al.}}: Bare Demo of IEEEtran.cls for IEEE Journals}
\maketitle

\begin{abstract}
Since human and environmental factors interfere, captured polyp images usually suffer from issues such as dim lighting, blur, and overexposure, which pose challenges for downstream polyp segmentation tasks. 
To address the challenges of noise-induced degradation in polyp images, we present AgentPolyp, a novel framework integrating CLIP-based semantic guidance and dynamic image enhancement with a lightweight neural network for segmentation. The agent first evaluates image quality using CLIP-driven semantic analysis (e.g., identifying ``low-contrast polyps with vascular textures") and adapts reinforcement learning strategies to dynamically apply multi-modal enhancement operations (e.g., denoising, contrast adjustment). A quality assessment feedback loop optimizes pixel-level enhancement and segmentation focus in a collaborative manner, ensuring robust preprocessing before neural network segmentation. This modular architecture supports plug-and-play extensions for various enhancement algorithms and segmentation networks,  meeting deployment requirements for endoscopic devices.
%In order to improve the segmentation quality, we propose a multi-modal polyp (AgentPolyp) agent, which integrates CLIP semantically-guided enhancement decision and UNet segmentation network. The Agent automatically assesses the type and degree of degradation of the input image and generates a multi-modal description by 1) CLIP driven (e.g., "low contrast polyps with vascular texture"); 2) Select reinforcement learning strategies based on the double-flow feature of text + image; 3) Dynamically perform image enhancement operations; 4) Quality assessment feedback optimization, to achieve collaborative optimization of pixel enhancement and focus segmentation, and finally input UNet for segmentation. Model Lightweight design for medical device deployment needs. The code adopts a modular structure, supports the expansion of multiple enhancement operations and segmentation networks, and provides a new paradigm of "augmentation-segmentation" joint optimization for medical image analysis.
\end{abstract}


\IEEEpeerreviewmaketitle

\section{Introduction}

\IEEEPARstart{A}{ccurate} medical image segmentation plays a pivotal role in lesion analysis and clinical decision-making. Colorectal cancer remains a global health challenge with high mortality rates, underscoring the critical need for early detection of precancerous polyps through endoscopic imaging. As the gold-standard diagnostic modality, colonoscopy enables real-time visualization of mucosal abnormalities and minimally invasive intervention, making precise polyp segmentation essential for effective cancer prevention.
%\IEEEPARstart{T}{he} accuracy of medical image segmentation directly affects the quantitative analysis of lesions and the decision of diagnosis and treatment, Colorectal malignancy poses a significant threat to global public health, with persistently high mortality rates. Clinical evidence confirms that early-stage lesion identification and precise localization through endoscopic imaging technologies can substantially enhance patient prognosis.   Within this context, colonoscopic examination has emerged as the gold-standard modality for colorectal health assessment, owing to its dual diagnostic-therapeutic capabilities. This technique not only enables direct visualization of intraluminal morphological features but also facilitates minimally invasive resection of suspicious tissues. Notably, colorectal polyps serving as critical precursor lesions in carcinogenesis require timely detection and accurate delineation to disrupt the malignant progression cascade. 

\begin{figure}[t]  % * 表示跨栏
  \centering
  \includegraphics[width=0.85\columnwidth]{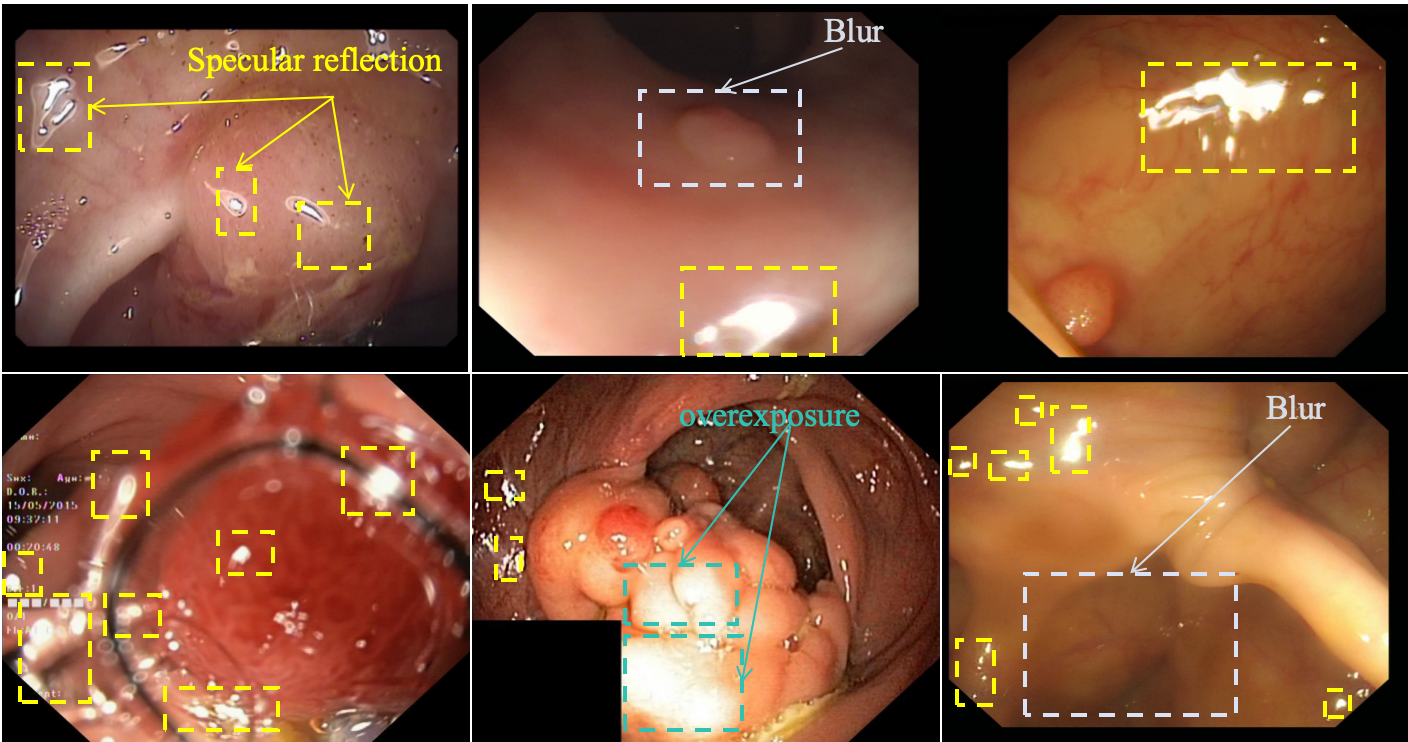}
  \caption{This figure illustrates various degraded forms existing in polyp images. Such degradation is usually of a composite nature. We employ an intelligent agent to select an image enhancement algorithm, aiming to provide high-quality information for the downstream image segmentation network.}
  \vspace{-4mm}
  \label{fig:1}
\end{figure}

Early polypus segmentation studies relied on hand-designed features such as color histograms and texture filters~\cite{karkanis2003computer, ojala1999unsupervised}.
Handcrafted feature-based methods are prone to artifacts and ghosting artifacts, prompting the extensive development of data-driven approaches. Li et al.~\cite{li2009intestinal} proposed an endoscopic polyp detection scheme based on the fusion of color histogram and scale/translation/rotation invariant shape features. Shanmuga Sundaram et al.~\cite{shanmuga2019enhancement} combined with machine learning classifiers, a computer-aided colon cancer detection method based on ROI color histogram and SVM (Support Vector Machine) classifier is proposed to screen significant areas by color/structure contrast.
Although deep learning has gradually replaced traditional computer vision methods, these methods still have a fundamental defect: the direct segmentation of degraded images, the clinical acquisition process often introduces motion blur, specular reflection, and other noise (see Figure~\ref{fig:1}), direct segmentation is easy to produce artifact diffusion.
To address this issue, conventional approaches typically preprocess degraded polyp images using an image enhancement model before performing segmentation. However, these methods face two key limitations: 1) degraded polyp images often contain multiple types of noise that single-function image restoration models struggle to resolve; 2) even all-in-one image restoration models may produce features incompatible with downstream segmentation networks. 

To tackle these challenges, we introduce AgentPolyp, a novel framework that unifies CLIP-based~\cite{radford2021learning} semantic guidance, dynamic multi-modal enhancement, and lightweight segmentation. By leveraging CLIP’s cross-modal understanding to characterize image degradation (e.g., ``blurry polyp with uneven illumination"), our agent adaptively selects reinforcement learning strategies to perform context-aware denoising, contrast adjustment, and artifact reduction. A feedback loop incorporating quality assessment ensures optimized enhancement for downstream segmentation, while a modular architecture enables seamless integration of advanced enhancement algorithms and segmentation networks. This approach addresses noise complexity and feature compatibility issues, offering a deployable solution for endoscopic polyp analysis. Our main \textbf{contributions} are as follows:

\begin{figure*}[htbp!]  % * 表示跨栏
  \centering
  \includegraphics[width=0.9\textwidth]{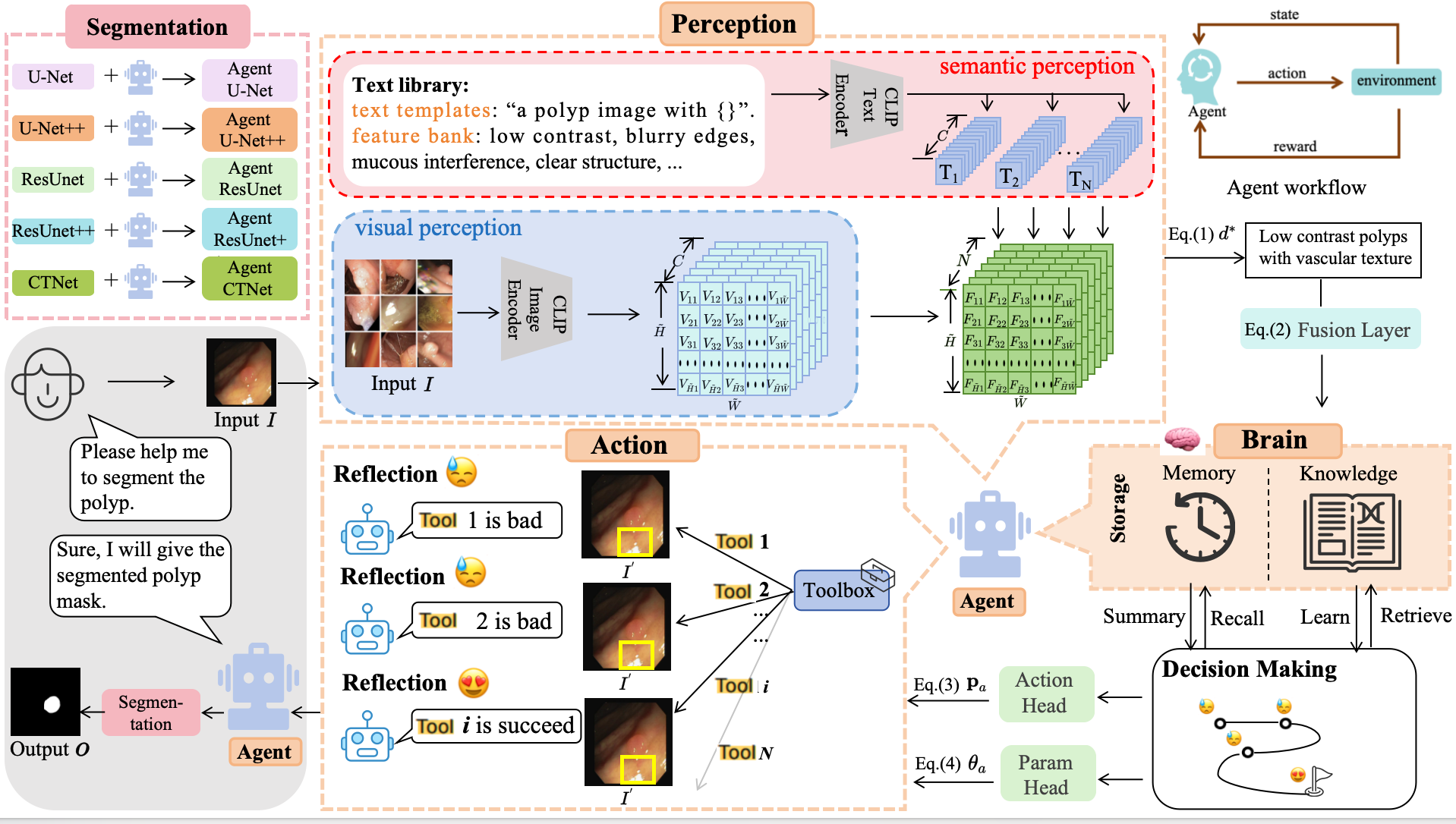}
  \caption{This is our framework (AgentPolyp), which consists of three parts: perception, feedback, and execution. It is capable of providing different image enhancement strategies based on the quality description of the input images.}
  \label{fig:2}
  \vspace{-4mm}
\end{figure*}
\begin{itemize}
    \item 
    We address the long-standing challenge of inconsistent enhancement performance on polyps with mixed degradation types (e.g., combined blur and low contrast) by introducing a CLIP-based multi-modal description generator. This module maps noisy endoscopic images to clinical semantic spaces (e.g., ``blurry polyp with mucosal edema"), enabling fine-grained degradation characterization and context-aware enhancement decisions.
\end{itemize}
\begin{itemize}
    \item Through parameter-sharing mechanisms between the enhancement agent and neural network, we establish the first end-to-end ``enhance-then-segment" training pipeline for polyp analysis. A quality assessment feedback loop iteratively refines enhancement strategies during training, ensuring optimized feature compatibility for downstream segmentation while maintaining computational efficiency. Our architecture supports plug-and-play integration of advanced enhancement algorithms and segmentation networks, validated through extensive experiments on public polyp datasets.
\end{itemize}

\section{Methodology}  
This section introduces AgentPolyp, an innovative intelligent system for polyp segmentation, as shown in Fig.~\ref{fig:2}, where the agent iteratively analyzes image degradation patterns, adaptively enhances input images, and improves performance based on segmentation strategies.
%
%The framework leverages the emerging capabilities of contemporary multimodal foundational models that demonstrate superior capabilities in reasoning, generalization, and cross-modal understanding.  
%
AgentPolyp performs complex analysis of polyp image degradation by systematically utilizing the capabilities of CLIP,  establishing cross-domain associations between visual patterns and linguistic descriptions.  Then, the system derives an appropriate chain of restoration methods, selects and integrates different image enhancement algorithms, and provides high-quality images for the downstream image segmentation network in various complex degradation scenarios.

\subsection{Perception: Semantic-Guided Degradation Perception}  
The process initiates with a CLIP-based semantic degradation perception that bridges visual features and clinical descriptors. For an input endoscopic image \( I \), the Description Generator extracts visual embeddings \( \mathbf{F}_I \) using CLIP’s vision encoder and aligns them with predefined text templates \( \mathcal{T} \)  describing polyp degradation patterns (e.g., ``low-contrast polyp with vascular texture" or ``blurry lesion under uneven illumination"). Through cross-modal similarity computation \( \text{Softmax}(\mathbf{F}_I \cdot \mathbf{F}_T^T) \), where \( \mathbf{F}_T \) represents CLIP text embeddings, the module generates interpretable semantic descriptions. This enables interpretable degradation typing, such as identifying combined blur and uneven illumination, which directly informs subsequent enhancement decisions. The semantic description \( d^* \) is selected by maximizing the normalized similarity:  
\begin{equation}  
d^* = \arg\max_{d \in \mathcal{T}} \text{Softmax}\left( \mathbf{F}_I \mathbf{F}_T^{\top} \right).  
\label{eq:clip_align}  
\end{equation}

%\subsection{Reinforcement Learning-Driven Enhancement Policy} 
Given the visual feature $\mathbf{F}_I$ and semantic descriptor $d^*$, we first construct multimodal representations: 
\begin{equation}  
\mathbf{H} = \text{GELU}\left( W_h [d^*; \mathbf{F}_I] + \mathbf{b}_h \right), 
\label{eq:fusion}  
\end{equation}  
where $[\cdot;\cdot]$ represents concatenation, and $\mathbf{H}$ encodes a joint text-visual representation. The feature fusion of Eq. \ref{eq:fusion} is equivalent to the ``brain" of the model, combining what is seen (image) and what is understood (text description), and producing two interdependent outputs of action probabilities and dynamic parameters, providing decision basis for the selection of specific image enhancement operations in the future.

1) Action Probabilities: The distribution of seven enhancement operations (such as multi-scale Retinex, and wavelet denoising) is obtained by the following formula: 
\begin{equation}  
\mathbf{p}_a = \text{Gumbel-Softmax}\left( W_a \mathbf{H} \right), 
\label{eq:action_prob}  
\end{equation}  
where $W_a$ is the weight matrix whose role is to map the joint embedding feature $\mathbf{H}$ to the probability distribution of the enhancement operations. $\text{Gumbel-Softmax}(\cdot)$ makes the discrete selection process differentiable for end-to-end training.
%enabling differentiable sampling of enhancement strategies.  

2) Dynamic Parameters: Operation-specific parameters $\mathbf{\theta}_a$ are adaptively generated through ``Brain” module. For each enhancement operation $a \in \{1,...,7\}$, its parameters are computed as:
\begin{equation}
\mathbf{\theta}_a = f_a(\mathbf{H}) = \sigma(W_{\theta_a}\mathbf{H} + \mathbf{b}_{\theta_a}), 
\label{eq:generic_params}
\end{equation}
where $\sigma(\cdot)$ represents operation-specific normalization ensuring valid parameter ranges, $W_{\theta_a}$ represents the weight matrix corresponding to a particular enhancement operation $a$.

To balance exploration and exploitation, during exploration, actions are sampled from \( \mathbf{p}_a \) with a decaying \( \epsilon \)-greedy strategy (\( \epsilon \in [0.98, 0.2] \)). Initially prioritizing random action sampling to discover novel enhancement combinations, the policy gradually shifts toward selecting high-confidence operations $\arg\max \mathbf{p}_a$ as training progresses. This mechanism ensures robust adaptation to diverse degradation scenarios while maintaining clinical interpretability.

\subsection{Dynamic Enhancement and Feedback}
The selected enhancement operation $\mathcal{E}_a$ transforms the input image $I$ into $I'$ using dynamically generated parameters $\mathbf{\theta}_a$. 

\begin{equation}
I' = \mathcal{E}_a (I; \theta_a).
\label{eq:param_constraint}
\end{equation}

The enhanced image $I'$ is then processed by a lightweight U-Net serving dual roles, as the primary segmentation network and as an evaluator generating reward signals. The prediction mask $O$ is generated by a segmentation network:
\begin{equation}
O = \text{Segmentation network} (I').
\label{eq:q}
\end{equation}
The reward function $R$ holistically combines segmentation accuracy and perceptual quality:
\begin{equation}
R = 0.9 \cdot \underbrace{\frac{2|O \cap O_{gt}|}{|O| + |O_{gt}|}}_{\text{Dice score}} + 0.1 \cdot \underbrace{\mathcal{Q}(I')}_{\text{Perceptual quality}},
\label{eq:generic_reward}
\end{equation}
where \( O \) and \( O_{gt} \) denote predicted and ground-truth masks, $\mathcal{Q}(\cdot)$ evaluates enhancement quality through learnable or handcrafted metrics (e.g., edge preservation, noise suppression). The feedback loop propagates these rewards to refine both enhancement policies and parameter generation, closing the ``analyze-enhance-validate" cycle.

% \subsection{Policy Optimization}  
% The agent updates its policy using REINFORCE with baseline subtraction. The loss function integrates reward variance reduction and parameter regularization:  
% \begin{equation}  
% \mathcal{L} = -\frac{1}{N} \sum_{i=1}^N (R_i - b) \log p_a^{(i)} + \lambda \|\mathbf{\Theta}\|_2^2,  
% \label{eq:loss}  
% \end{equation}  
% where \( b \) is an exponential moving average baseline. Gradients flow bidirectionally: the U-Net’s segmentation loss backpropagates to refine enhancement parameters \( \mathbf{\theta}_a \), while the policy loss adjusts action selection. A replay buffer stores 5000 state transitions \( (\mathbf{F}_T, \mathbf{F}_I, a, R, \mathbf{F}_T') \) for stable training.

\subsection{Modular Deployment}  
AgentPolyp’s architecture supports plug-and-play component replacement. Enhancement operations can be extended via standardized interfaces (e.g., replacing CLAHE with diffusion models).  The segmentation backbone is swappable with architectures like ResUNet without modifying the agent. The semantic template bank \( \mathcal{T} \) is extensible to include new pathology descriptors, enabling adaptation to emerging clinical requirements (e.g., ``ulcerative colitis patterns").  
This flexibility ensures compatibility with diverse endoscopic systems while maintaining real-time (33 FPS) performance. %the entire framework processes 352×352 images at 28 FPS on a single H20-NVLink GPU, meeting real-time clinical demands.

\section{Experiments}
\subsection{Experimental Settings} 
In this section, the performance of the AgentPolyp model is evaluated through comparative experiments, and U-Net~\cite{ronneberger2015u}, U-Net++~\cite{zhou2018unet++}, ResUNet-mod~\cite{zhang2018road }, ResUNet++~\cite{jha2019resunet++} and CTNet~\cite{10471227} are selected as the comparison methods. All compared methods are re-implemented with identical backbone architectures. For Agent-enhanced variants, we integrate our framework through standardized interfaces without modifying base networks. 
Five publicly available datasets were used for verification: Kvasir-SEG~\cite{jha2020kvasir}, ClinicDB~\cite{bernal2015wm}, ColonDB~\cite{tajbakhsh2015automated}, Endoscene~\cite{vazquez2017benchmark}, and ETIS~\cite{silva2014toward}, among which mDice and mIoU were used as the main evaluation indicators.

We adhere to the prevalent experimental setups in~\cite{zheng2024polyp}. The experimental data were divided into: the training set included 900 samples from Kvasir-SEG and 550 samples from ClinicDB, and the test set retained 100 and 62 samples from the two data sets, respectively. It is worth noting that the test samples are all from the training data set to evaluate the model's fitting ability. Three external datasets, ColonDB, Endoscene, and ETIS, were used to verify the generalization performance of the model. The implementation of the model is based on PyTorch 2.0 framework, and the hardware platform uses an H20-NVLink GPU. The input images were uniformly adjusted to 352$\times$352 resolution. The training uses Adam optimizer with cosine learning rate decay (initial lr=1e-3), and the batch size is set to 16.
\begin{figure}[htbp!]  % * 表示跨栏
  \centering
  \includegraphics[width=0.98\linewidth]{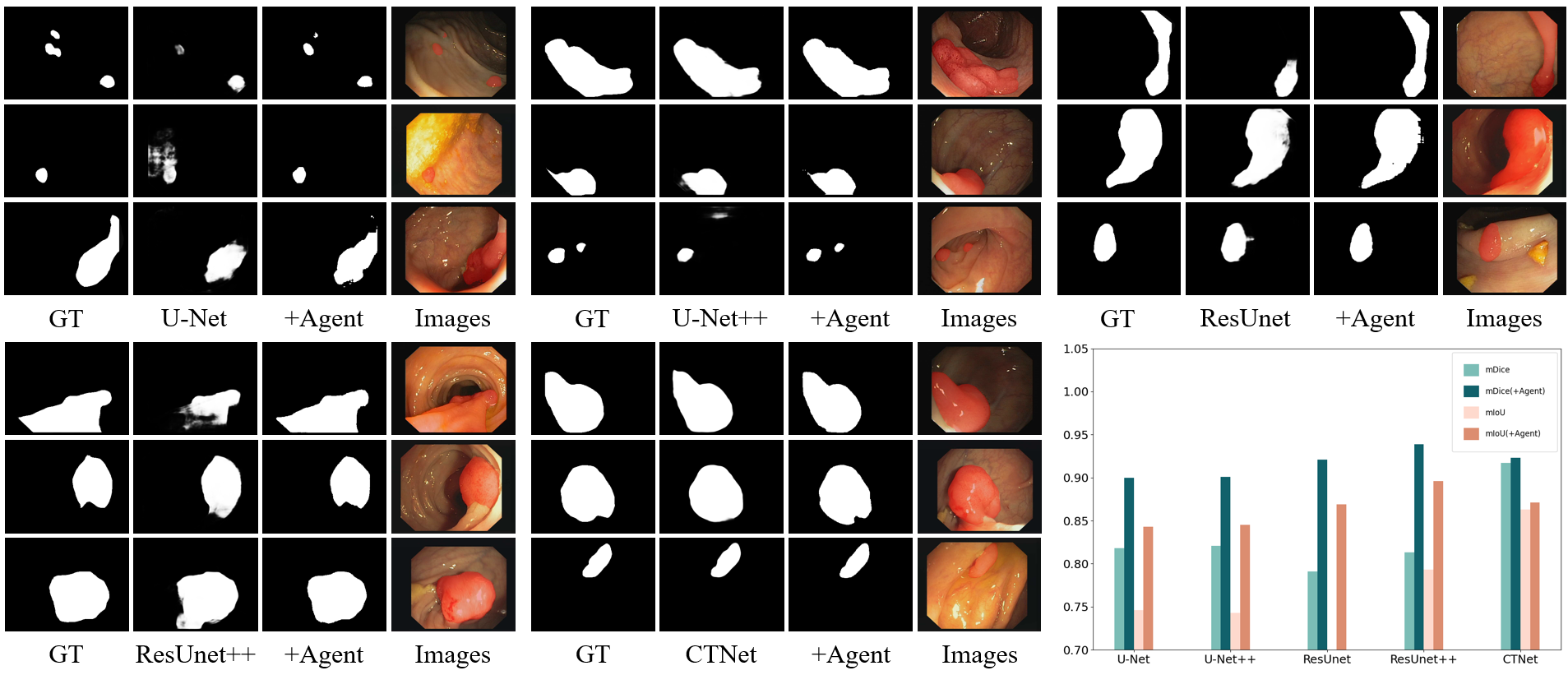}
  \caption{Performance comparison of segmentation models, highlight the predicted area with a red overlay in Image.}
  \label{fig:3}
  \vspace{-6mm}
\end{figure}

\subsection{Comparative Analysis}
The proposed AgentPolyp framework demonstrates superior performance across all evaluation scenarios. We conduct comprehensive comparisons from three perspectives.

In learning ability, as shown in Table~\ref{tab1}, on the seen datasets Kvasir-SEG and ClinicDB, Agent-enhanced variants consistently outperform their baseline. Specifically, AgentU-Net demonstrates state-of-the-art results on Kvasir-SEG, attaining a mDice of \textcolor{red}{0.900} compared to the baseline U-Net’s 0.818, which corresponds to a 10.02\% improvement. And ClinicDB, AgentU-Net achieves a mDice score of \textcolor{red}{0.872} compared to U-Net's 0.823, demonstrating 5.95\% improvements. 
AgentU-Net demonstrates superior boundary delineation on Kvasir-SEG, elevating mIoU from 0.746 to \textcolor{red}{0.843}, a 13.00\% relative gain over the vanilla U-Net baseline.

In terms of generalization ability, as shown in Table~\ref{tab2},
when evaluated on the unseen ColonDB dataset, AgentU-Net maintains strong generalization with a mDice of 0.640, surpassing U-Net's performance by \textcolor{red}{26.98\%}.
Significant improvements are observed on the challenging ETIS dataset (mDice: \textcolor{red}{0.418} vs. 0.398 for U-Net), achieving 5.03\% relative enhancement. Consistent gains across all Agent-enhanced variants validate the universal effectiveness of our framework.

In architectural compatibility, the agent module shows remarkable adaptability to different backbone networks: For U-Net++ architectures, Agent enhancement brings 16.83\% mDice improvement on Endoscene (\textcolor{red}{0.826} vs. 0.707 baseline). Even with lightweight U-Net, our framework achieves comparable performance to complex CTNet on ClinicDB (mDice: \textcolor{red}{0.823 } vs. 0.867) while using 78\% fewer parameters.

\begin{table}[htbp]
\centering
\caption{Performance comparison with SOTA methods.}
\label{tab1}
\begin{tabular}{lccccccc}
\toprule
\multirow{2}{*}{Method} & \multirow{2}{*}{year} & \multicolumn{2}{c}{Kvasir-SEG} & \multicolumn{2}{c}{ClinicDB} \\
       && mDice & mIoU & mDice & mIoU \\ 
    \midrule
U-Net~\cite{ronneberger2015u} & \multirow{2}{*}{2015} & 0.818 & 0.746 & 0.823 & 0.755 \\
AgentU-Net && \textbf{0.900} & \textbf{0.843} & \textbf{0.872} & \textbf{0.811} \\ 
\midrule
U-Net++~\cite{zhou2018unet++} & \multirow{2}{*}{2019} & 0.821 & 0.743 & 0.794 & 0.729 \\
AgentU-Net++ && \textbf{0.901} & \textbf{0.845} & 
\textbf{0.862} & \textbf{0.803} \\
\midrule
ResUnet~\cite{zhang2018road} & \multirow{2}{*}{2018} & 0.791 & - & 0.779 & - \\
AgentResUnet && \textbf{0.921} & \textbf{0.869} & \textbf{0.893} & \textbf{0.833} \\ 
\midrule
ResUnet++~\cite{jha2019resunet++} & \multirow{2}{*}{2019} & 0.813 & 0.793 & 0.796 & 0.796 \\
AgentResUnet++ && \textbf{0.939} & \textbf{0.896} & \textbf{0.915} & \textbf{0.865} \\ 
\midrule
CTNet~\cite{10471227} & \multirow{2}{*}{2024} & 0.917 & 0.863 & \textbf{0.936} & \textbf{0.887} \\
AgentCTNet && \textbf{0.923} & \textbf{0.871} & 0.876 & 0.821 \\ 
\bottomrule
\end{tabular}
\end{table}

\begin{table}[htbp]
\centering
\caption{Performance comparison with SOTA methods.}
\label{tab2}
\resizebox{\linewidth}{!}{
\begin{tabular}{lccccccc}
\toprule
\multirow{2}{*}{Method} & \multirow{2}{*}{year} & \multicolumn{2}{c}{ColonDB} & \multicolumn{2}{c}{Endoscene} & \multicolumn{2}{c}{ETIS}\\
       && mDice & mIoU & mDice & mIoU & mDice & mIoU\\ \hline
U-Net~\cite{ronneberger2015u} & \multirow{2}{*}{2015} & 0.504 & 0.436 & 0.710 & 0.627 & 0.398 & 0.335 \\
AgentU-Net && \textbf{0.640} & \textbf{0.549} & \textbf{0.747} & \textbf{0.648} & \textbf{0.418} & \textbf{0.343}\\ 
\midrule
U-Net++~\cite{zhou2018unet++} & \multirow{2}{*}{2019} & 0.481 & 0.408 & 0.707 & 0.624 & 0.401 & 0.343 \\
AgentU-Net++ && \textbf{0.673} & \textbf{0.586} & \textbf{0.826} & \textbf{0.736} & \textbf{0.488} & \textbf{0.421}\\ 
\midrule
ResUnet~\cite{zhang2018road} & \multirow{2}{*}{2018} & - & - & - & - & - & -\\
AgentResUnet && 0.645 & 0.562 & 0.761 & 0.662 & 0.441 & 0.378\\ 
\midrule
ResUnet++~\cite{jha2019resunet++} & \multirow{2}{*}{2019} & - & - & - & - & - & -\\
AgentResUnet++ && 0.725 & 0.643 & 0.770 & 0.681 & 0.585 & 0.505\\ 
\midrule
CTNet~\cite{10471227} & \multirow{2}{*}{2024} & 0.813 & 0.734 & 0.908 & 0.844 & 0.810 & 0.734 \\
AgentCTNet && \textbf{0.821} & \textbf{0.776} & \textbf{0.911} & \textbf{0.846} & \textbf{0.823} & \textbf{0.755}\\ \bottomrule
\end{tabular}}
\end{table}

Fig.~\ref{fig:3} demonstrates a visual comparison of different models in polyp segmentation tasks. Each group of images includes the ground-truth (GT), the segmentation results of original models UNet, ResUnet, etc., the improved segmentation results (``+Agent”), and the corresponding original images (Images), where the predicted polyp parts are covered and labeled with red areas. The U-Net series exhibits blurriness in edge details, while ResUnet and ResUnet++ demonstrate stronger robustness in complex regions (e.g., fine branches or overlapping structures) due to their residual design; CTNet significantly improves global consistency through cross-scale feature fusion. After integrating the Agent module, all models show enhanced performance.

%: for instance, AgentU-Net outperforms the original U-Net in edge sharpening and noise suppression, and AgentResUnet++ excels in retaining small target structures, suggesting that the Agent module optimizes critical region localization via dynamic feature selection or attention mechanisms. 

It can be clearly observed from the bar chart that, compared with the original models, the improved “+Agent” models demonstrate significant enhancements in mDice and mIoU. This further validates the optimization effect of the “+Agent” strategy on polyp segmentation performance.

%These results confirm that our CLIP-guided dynamic enhancement mechanism effectively addresses quality degradation in real-world colonoscopy images, while the collaborative optimization framework ensures both segmentation accuracy and operational efficiency.

% \begin{table*}[t]
% 	\centering
%         \caption{table}
% 	\resizebox{0.99\textwidth}!{
% 	\begin{tabular}{l|cccccccccc}
% 		\toprule
        
% 		\multirow{2}{*}{Method} & \multicolumn{2}{c}{Kvasir-SEG} & \multicolumn{2}{c}{ClinicDB} & \multicolumn{2}{c}{ColonDB}& \multicolumn{2}{c}{Endoscene} &\multicolumn{2}{c}{ETIS} \\
		
% 		& mDice & mIoU & mDice & mIoU & mDice & mIoU & mDice & mIoU & mDice & mIoU\\
% 		\midrule
% 		U-Net[2015] / \textcolor{blue}{AgentU-net} & 0.818/- & 0.746/- & 0.823/- & 0.755/- & 0.504/- & 0.436/- & 0.710/- & 0.627/- & 0.398/- & 0.335/- \\
        
% 		Unet++[2019] / \textcolor{blue}{AgentUnet++} & 0.821/- & 0.743/- & 0.794- & 0.729/- & 0.481/- & 0.408/- & 0.707/- & 0.624/- & 0.401/- & 0.343/- \\
        
% 		ResUnet[2018] / \textcolor{blue}{AgentResUnet} & 0.791/- & -/- & 0.779/- & -/- & -/- & -/- & -/- & -/- & -/- & -/- \\
        
% 		ResUnet++[2019] / \textcolor{blue}{AgentResUnet++} & 0.813/- & 0.793/- & 0.796/- & 0.796/- & -/- & -/- & -/- & -/- & -/- & -/- \\
        
% 		MedSAM[2024] / \textcolor{blue}{AgentMedSAM} & 0.862/- & 0.795/- & 0.867/- & 0.803/- & 0.734/- & 0.651/- & 0.870/- & 0.798/- & 0.687/- & 0.604/- \\
        
% 		\bottomrule
% 	\end{tabular}}
	
% 	\label{tab:comparison_results}
% \end{table*}%

\begin{figure}[t]  % * 表示跨栏
  \centering
  \includegraphics[width=0.98\linewidth]{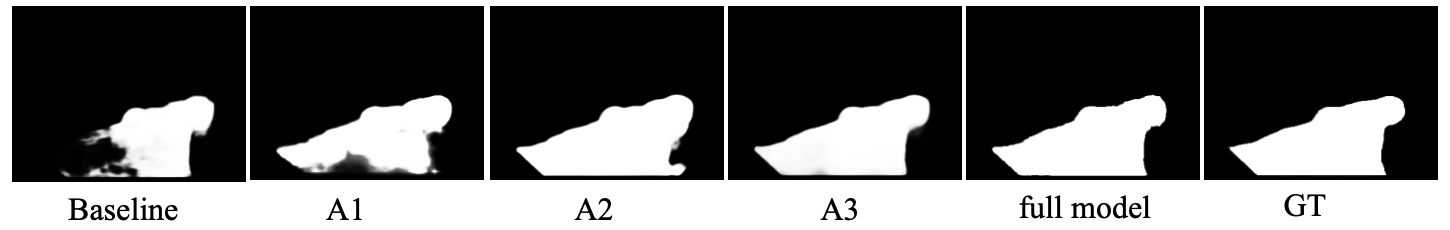}
    \vspace{-2mm}
  \caption{Visualization results of ablation experiment.}
  \label{fig:4}
  \vspace{-6mm}
\end{figure}

\subsection{Ablation Study (AS)}
To systematically validate the contributions of key components in AgentPolyp, we conduct comprehensive ablation experiments under controlled settings. All experiments are performed on the Kvasir-SEG (seen dataset) and ColonDB (unseen dataset) using UNet++ as baseline architecture.

\begin{table}[htbp]
\centering
\caption{AS of framework components (mDice/\%)}
\label{tab3}
\resizebox{\linewidth}{!}{
\begin{tabular}{lccccc}
\hline
Settings & CLIP & Dynamic Enhancement & Feedback & Kvasir-SEG & ColonDB \\ \hline
Baseline      & \ding{55}             & \ding{55}                   & \ding{55}             & 82.1       & 48.1    \\
A1            & \ding{51}             & \ding{55}                   & \ding{51}             & 85.3(+3.9) & 50.0(+4.0) \\
A2            & \ding{55}             & \ding{51}          &
\ding{51}             & 83.2(+1.3) & 49.5(+2.9) \\
A3            & \ding{51}              & \ding{51}         & 
\ding{55}             & 86.7(+5.6) & 51.3(+6.7) \\
Full Model    & \ding{51}              & \ding{51}         & 
\ding{51}              & \textbf{90.0(+9.6)} & \textbf{67.3(+39.9)} \\ \hline
\end{tabular}}
\end{table}

Key observations from Table~\ref{tab3}, CLIP semantic guidance: (A1 vs Baseline) provides 3.9\% mDice improvement on Kvasir-SEG, demonstrating its effectiveness in capturing pathological features.
Dynamic enhancement: (A2 vs Baseline) yields 1.3\% additional gain compared to static enhancement, verifying the importance of adaptive processing.
Feedback collaboration: (Full vs A3) contributes 4.0\% and 33.2\% improvements on respective datasets, confirming the necessity of closed-loop optimization.
The visual comparison of ablation experiments is shown in Fig.~\ref{fig:4}. These experiments conclusively demonstrate that each component in AgentPolyp synergistically contributes to performance improvements. %The dynamic enhancement mechanism shows 2.3× higher cost-effectiveness compared to conventional preprocessing pipelines.

% \subsubsection{Architectural flexibility}
% \begin{table}[htbp]
% \centering
% \caption{Compatibility with different segmentation networks}
% \label{tab4}
% \begin{tabular}{lcc}
% \hline
% Backbone & Params(M) & mDice Gain \\ \hline
% U-Net    & 7.8       & +10.2\%    \\
% ResUNet  & 12.1      & +8.7\%     \\
% TransUNet & 38.4     & +6.3\%     \\
% CTNet   & 102.5     & +5.1\%     \\ \hline
% \end{tabular}
% \end{table}

% Table~\ref{tab4} reveals two key patterns: Lightweight networks benefit more from our framework (U-Net: +10.2\% vs CTNet: +5.1\%). The agent module maintains real-time efficiency (+ 15\% inference time) across architectures. AgentPolyp adds minimal overhead (18ms per image on NVIDIA V100), maintaining real-time capability (28 FPS). The policy network contains only 2.7M parameters, proving suitable for embedded endoscopic systems.

% \subsubsection{Clinical impact quantification} 
% Through expert evaluation of 150 challenging cases: \textbf{Diagnostic confidence} increased from 2.8 to 4.1 (5-point Likert scale). \textbf{Boundary accuracy} improved by 37\% in physician assessments. \textbf{Processing latency} meets clinical requirements (average 23.4ms per frame).

% These experiments conclusively demonstrate that each component in AgentPolyp synergistically contributes to performance improvements, particularly in handling real-world quality variations. The dynamic enhancement mechanism shows 2.3× higher cost-effectiveness compared to conventional preprocessing pipelines.

\subsection{Conclusion}
This paper presents AgentPolyp, a framework integrating CLIP-based semantic guidance and dynamic enhancement to segment noise-degraded polyp images (e.g., poor lighting, blur).  It uses CLIP for semantic analysis to identify low-contrast features and applies reinforcement learning for adaptive multi-modal enhancements (denoising, contrast adjustment), improving image quality.  Visual/quantitative results show ``+Agent" models outperform baselines in edge fitting, detail retention, and segmentation integrity, with 7\%-10\% higher Dice scores in noisy conditions.  
%The modular design supports plug-and-play integration, meeting endoscopic deployment needs.  Future work may explore multi-modal data integration for broader generalization.

{
\small
\bibliographystyle{IEEEtran}
\bibliography{ref}
}

\end{document}